\documentclass[showpacs,twocolumn,aps,superscriptaddress,preprintnumbers,nofootinbib]{revtex4}

\usepackage{amsmath,amssymb}
\usepackage{epsfig}
\usepackage{amsmath}
\usepackage{amsfonts}
\usepackage{epstopdf}
\usepackage{color}

\newcommand{\be}{\begin{equation}}
\newcommand{\ee}{\end{equation}}
\newcommand{\bear}{\begin{eqnarray}}
\newcommand{\eear}{\end{eqnarray}}

\newcommand{\ba}{\begin{array}}
\newcommand{\ea}{\end{array}}

\begin{document}
\preprint{OU-HET-911}

\title{Universality in Chaos of Particle Motion near Black Hole Horizon}

\author{Koji Hashimoto}
\author{Norihiro Tanahashi}
\affiliation{Department of Physics, Osaka University, Toyonaka, Osaka 560-0043, Japan}

\begin{abstract} 
Motion of a particle near a horizon of a spherically symmetric black hole is shown to possess a universal Lyapunov exponent
of a chaos provided by its surface gravity. To probe the horizon, we introduce electromagnetic or scalar
force to the particle so that it does not fall into the horizon. There appears an unstable maximum of the
total potential where the evaluated maximal Lyapunov exponent is found 
to be independent of the external forces and the particle mass.
The Lyapunov exponent is universally given by the surface gravity of the black hole. 
Unless there are other sources of a chaos,
the Lyapunov exponent is subject to
an inequality 
$\lambda \leq 2\pi T_{\rm BH}/\hbar$, which is identical to the bound recently discovered by Maldacena, Shenker and Stanford.
\end{abstract}


\maketitle

\setcounter{footnote}{0}



\section{Introduction.}

Recent discovery of a universal upper bound of a Lyapunov exponent
defined in quantum field theories, provided by Maldacena, Shenker and Stanford
\cite{Maldacena:2015waa}, 
attracts much attention, as it can bridge physics of black holes and quantum information.
The Lyapunov exponent $\lambda$ of so-called out-of-time-ordered correlators \cite{Larkin,Shenker:2013pqa}
of the theories is expected by the temperature $T$,
\begin{align}
\lambda \leq \frac{2\pi T}{\hbar} \, .
\label{MSS}
\end{align}
The characteristic $T$-dependence was originally derived by thought-experiments of shock waves near
black hole horizons \cite{Shenker:2013pqa,Shenker:2013yza}
(and subsequent papers \cite{Leichenauer:2014nxa,Kitaev-talk,Shenker:2014cwa,Jackson:2014nla,Polchinski:2015cea}) and 
the AdS/CFT correspondence \cite{Maldacena:1997re}.
The bound is saturated by the Sachdev-Ye-Kitaev model \cite{Sachdev:1992fk,Kitaev-talk-KITP}, which
ignited interesting development (see for example \cite{Sachdev:2015efa,Hosur:2015ylk,Gur-Ari:2015rcq,Stanford:2015owe,Polchinski:2016xgd,Michel:2016kwn,Anninos:2016szt,Turiaci:2016cvo,Jevicki:2016bwu,Maldacena:2016hyu,Jensen:2016pah,Maldacena:2016upp,Engelsoy:2016xyb,Grumiller:2016dbn,Brown:2016wib,Hartnoll:2016mdv,Cvetic:2016eiv,Jevicki:2016ito,Giombi:2016pvg,Polchinski:2016hrw}, and also \cite{Hartman:2015lfa,Brown:2015bva,Berenstein:2015yxu,Fitzpatrick:2016thx,Perlmutter:2016pkf,Fu:2016yrv,Blake:2016wvh,Fitzpatrick:2016ive,Roberts:2016wdl,Blake:2016sud,Hashimoto:2016wme,Danshita:2016xbo,Miyaji:2016fse,Halpern:2016zcm}). 
One of the important questions is, among various physics associated with black hole horizons,
what provides the bound, other than the shock waves.

In history, chaos of motion of a relativistic particle has been 
extensively studied. Classic examples include various motion of a particle
orbiting around black holes (for example, see \cite{Suzuki:1996gm,Sota:1995ms} and references therein).
In particular,
it would leave an imprint on gravitational wave emitted from it~\cite{Suzuki:1999si,Kiuchi:2004bv,Kiuchi:2007ue}.
Observation of gravitational wave, which was recently realized by LIGO experiment \cite{Abbott:2016blz,Abbott:2016nmj}, might be able to detect such chaotic behavior around black holes.
There were also an attempt to relate unstable circular geodesics of null rays and their Lyapunov exponents to quasi-normal mode frequencies~\cite{Cardoso:2008bp}.%
\footnote{See also Ref.~\cite{Gibbons:2008hb,Barbon:2011pn,Barbon:2011nj,Barbon:2012zv} for studies on optical geometries and their relation to chaos associated to black hole horizons.}
However, the universal feature of the Lyapunov exponent is yet to be uncovered.

In this paper, we aim to
relate these two subjects which have developed independently.
We consider a relativistic particle moving near a black hole horizon. It is not orbiting around
the black hole, but pulled outward by some external forces such as electromagnetic or scalar
forces. The force is so strong that the particle can come very close to the horizon
without falling into it.
As a result, we find a local maximum of the total effective potential
at which a Lyapunov exponent $\lambda$
for particle trajectories
can be evaluated. Surprisingly, we find that the exponent
is independent of the strength and the species of the external force, and also 
of the particle mass. It is given by
\begin{align}
\lambda = \kappa\, ,
\label{lyapuk}
\end{align}
where $\kappa$ is the surface gravity of the black hole horizon.
The maximum of the potential works as a separatrix of the phase space motion of the
particle, therefore, 
unless there are some other sources of the chaos, 
the evaluated Lyapunov exponent is an upper bound of Lyapunov exponents
of chaotic motion of the particle: the horizon 
is a nest of the chaos. Since the surface gravity is equal to
$2\pi T/\hbar$ where $T$ is the Hawking temperature of the black hole,
our universal upper bound is identical to \eqref{MSS}.

The paper is organized as follows. First in section \ref{sec:PM}, we derive 
effective Lagrangian of a particle pulled by electromagnetic or scalar force,
near a black hole horizon.
In section \ref{sec:nest}, we evaluate the Lyapunov exponent at the separatrix
and describe the universality.
In section \ref{sec:num}, we perform numerical simulation of the particle motion
and demonstrate the existence of the chaos, with a brief evaluation of the Lyapunov exponent.
In section \ref{sec:ang}, we revisit a model of orbiting neutral particle and find that
it cannot come closer to the horizon to realize \eqref{lyapuk}.
In section \ref{App:generic}, we analyze generic potentials and show that
higher spin forces violate the bound.
The final section is for our conclusion.
In appendices, after reviewing the surface gravity and temperature of a black hole horizon in appendix~\ref{App:surfacegravity},
we provide a simplified model with a chaos, to discuss
a relation between the chaos and redshift in appendix~\ref{sec:Red}.
We also examine the particle motion with scalar force taking the gravitational backreaction into account in appendix~\ref{App:scalar}.

\section{Particle motion near black hole horizons.}
\label{sec:PM}

The theme of the study is to show a universal behavior appearing in the motion
of a relativistic particle near black hole horizons. In this section we derive
effective Lagrangians of a particle which can come very close to the horizon.

Let us consider a relativistic particle with mass $m$ in $d$ spacetime dimensions.
We are particularly interested in it moving near a 
black hole horizon. Let us consider a black hole with a metric given by
\begin{equation}
 ds^2 = g_{tt}(r) dt^2 + g_{rr}(r) dr^2 + \cdots.
\end{equation}
At the horizon, the factor $g_{rr}$ diverges, and
the surface gravity $\kappa$ of this black hole (see Appendix A) is
given by 
\begin{align}
\kappa=\sqrt{\frac{-g_{tt}}{g_{rr}}}
 \frac{d}{dr}\left(\log\sqrt{-g_{tt}}\right)\biggm|_{\rm horizon} \, .
 \label{surfacegravity}
\end{align}
We consider spherically symmetric and static black holes given by
\begin{equation}
 ds^2 = -f(r) dt^2 + \frac{dr^2}{g(r)} + r^2 d\Omega_{d-2} \, .
 \label{smet}
\end{equation}
At the horizon, the function $f(r)$ and $g(r)$ are expanded as
\begin{align}
f(r) = \alpha_f (r-r_g)^{\beta_f},
\qquad
g(r) = \alpha_g (r-r_g)^{\beta_g}.
\label{expandg}
\end{align}
A non-extreme regular black hole satisfies $\beta_f = \beta_g = 1$, and $\beta_f = \beta_g = 2$ corresponds to a usual extreme black hole.%
\footnote{%
See \cite{Visser:1992qh,Medved:2004ih,Medved:2004tp,Tanatarov:2012xj} for allowed parameter range in more general situations.}
In these cases, the surface gravity is calculated as
\begin{align}
\kappa
&=
\frac12 p_f \sqrt{\alpha_f \alpha_g} (r-r_g)^{\frac12\left(\beta_f + \beta_g - 2\right)}
\biggm|_{r=r_g} 
\notag \\
&= 
\begin{cases}
\frac12 \sqrt{\alpha_f \alpha_g}
& 
\!\!
\mathrm{for} ~ \beta_f = \beta_g = 1
\\
0 & 
\!\!
\mathrm{for}~\beta_f = \beta_g = 2
\end{cases}
~.
\end{align}
These expressions are valid also for spherically symmetric black holes in
asymptotically AdS spaces and for large black holes.

The Lagrangian of a particle with mass $m$ moving in an external potential $V(x^\mu)$ is given by
\begin{align}
{\cal L} = -m \sqrt{- g_{\mu\nu}(X)\dot{X}^\mu \dot{X}^\nu} - V(X) \, .
\end{align}
Here $g_{\mu\nu}$ is the metric with 
$\mu,\nu=0,\cdots, d-1$. We take a static gauge $X^0=t$, then
in the given metric \eqref{smet}, the Lagrangian is
\begin{equation}
 {\cal L} = - m\sqrt{
f(r) - \frac{1}{g(r)}\dot r^2 - r^2 \dot \theta^2
}
- V(r,\theta) \, .
\end{equation}
This expression assumes that the particle moves only in the $r, \theta$ directions.
The dependence on other angular variables can be incorporated trivially.

Our focus is on the near-horizon region of the black holes with
$\beta_f = \beta_g = 1$,
the standard case.
Using the expansion \eqref{expandg} with the distance from the horizon denoted by 
$x\equiv r-r_g$, 
the particle Lagrangian is written as
\begin{gather}
  {\cal L} = - m\sqrt{
 \alpha_f x - \frac{1}{\alpha_g x}\dot x^2 - \dot y^2
  }
  - V(x,y) \, ,
\end{gather}
where we introduced a coordinate $y \equiv r_g \theta$.
To focus on slow motion of the particle,
we 
take the non-relativistic limit by assuming the time-derivative terms are sufficiently smaller than the other part.
If the potential $V(r)$ in these coordinates is regular at the horizon, we may generically 
be able to approximate the external potential as a linear one along the radial $x$ direction,
\begin{align}
V(r) = c m  x \, ,
\label{linearV}
\end{align}
where $c$ is some negative coefficient,
for which the Lagrangian reduces to
\begin{equation}
 {\cal L}/m =
  - \sqrt{\alpha_f x}
  \left[
   1 - \frac{\dot x^2}{2 \alpha_f \alpha_g x^2}
  \right]
  - c x  \, .
  \label{Lm}
\end{equation}
The angular dependence ($y$-dependence) is not important and we omit it here.
The potential $V$ for the $y$ direction can be flat or even harmonic $\sim y^2$,
and we assume that the $y$ motion does not contribute much to the following argument.
The angular motion of a particle will be discussed in Sec.~\ref{sec:ang} independently.

\begin{figure}
\includegraphics[width=6cm]{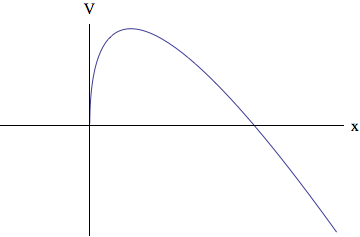}
\caption{A schematic plot of the effective potential $V_{\rm eff}$ given in \eqref{Veff}. The origin of
the plot is the horizon $x=0$. The potential has a maximum near the horizon, due to the
competition between the gravitational potential and the external force.}
\label{fig:pot}
\end{figure}

From Eq.~\eqref{Lm}, we can define the effective potential $V_\text{eff}$ by 
\begin{equation}
 V_\text{eff}  = \sqrt{\alpha_f x} + c x \, .
  \label{Veff}
\end{equation}
The first term is the gravitational potential which behaves as $\sqrt{x}$ near the horizon $x=0$.
The second term is the external potential which we introduced in order to control the particle probing the horizon.
Since we need the information back, the particle should not fall into the horizon,
so the coefficient $c$ is taken to be negative.
With this combination one can easily find that the potential has an unstable extremum
near the horizon. See Fig.~\ref{fig:pot}. The position of the extremum is given by
\begin{align}
 x_0 = \frac{\alpha_f}{4c^2} \, .
\end{align}
If we increase the energy of the particle, at some energy the
particle can reach the extremum. 
The particle motion around the extremum is slow enough, and the non-relativistic approximation is 
valid around there. Around the top of the potential, we can expand the effective potential
as
\begin{equation}
 V_\text{eff}(r)=
  - \frac{\alpha_f}{4c}
  - \frac{\sqrt{\alpha_f}}{8x_0^{3/2} } 
  (x-x_0)^2
  + \cdots.
\end{equation}
Then, around $x = x_0$, the $x$ part of the Lagrangian is approximated by
\begin{equation}
 {\cal L} \sim
  - \frac{mc^3}{2\kappa^3}
  \sqrt{\frac{\alpha_g}{\alpha_f}}
  \left[
   \dot x^2 + \kappa ^2 (x-x_0)^2
  \right].
  \label{Lag}
\end{equation}
This is an inverse harmonic oscillator with a specific background dependence.
A notable feature of this Lagrangian is that the slope of the potential $c$ contributes only to the overall factor, 
and so does the mass $m$ and the ratio of the metric components $\alpha_g/\alpha_f$ ---
the particle dynamics is sorely characterized by $\kappa$.
In the next section, we study how this action can characterize the chaos around the black hole horizon.

In deriving the particle motion given by the Lagrangian \eqref{Lag},
we have assumed that the potential $V$ can be approximated by a linear function. 
We can validate this assumption by adopting a standard electric force to generate the potential $V$. 
 The electrostatic potential $A_0$ in the vacuum near the horizon 
is governed by a field equation
\begin{align}
\partial_r \left(
\sqrt{-\det g} \, g^{rr} g^{00} \partial_r A_0 
\right)= 0 \, .
\label{eleE}
\end{align}
This particular metric dependence is important for us, since the factor 
$\sqrt{-\det g} \, g^{rr} g^{00} $ does not have any singular behavior at the horizon.
In fact it is merely a nonzero constant at the leading order
of the near-horizon expansion.
Therefore, the electrostatic potential $A_0$ is solved by a linear function of $r$
near the horizon.
Supposing that the point particle has a charge $e$, the potential term in the action
can be written simply in the present coordinate system, in a reparameterization-invariant and gauge-invariant manner, as
\begin{align}
V(X) = e \frac{dX^0}{dt} A_0(X) \, ,
\end{align}
which is equal to $e A_0(X)$ with our gauge fixing $X^0=t$. So, the potential term
$V$ in the Lagrangian \eqref{Lag} is equal to the electrostatic potential
which is a linear function.
This validates our assumption to derive the particle action \eqref{Lag}.

If the electric field is created by the black hole itself, the spherically symmetric black hole is
a Reissner-Nordstr\"om black hole. It has spherically symmetric Coulomb potential $A_0$,
which is linear in $x$ if it is expanded around any point on the horizon. So, the
Reissner-Nordstr\"om black hole is one example to realize our set-up. In general, the 
electric field could be prepared by any other sources.

Other candidate for the potential is a scalar force. Here we show that
even with the scalar force the resultant particle action near the extremum of the potential has the same form as \eqref{Lag}.\footnote{
Normally, the black hole horizon develops
a singularity in the presence of nontrivial scalar field (see \cite{Herdeiro:2015waa} 
for a recent review on the scalar hair), thus scalar
force is not suitable for our purpose
unless some mechanism is introduced to avoid singularity formation due to gravitational backreaction.
Here we simply ignore the
backreaction of the scalar field onto the metric.
We discuss the effect of the backreaction in Appendix~\ref{App:scalar}.}
In the presence of the scalar potential $\phi(x)$, a reparameterization-invariant
Lagrangian of a particle is
\begin{align}
{\cal L}= -\sqrt{-g_{\mu\nu}(X)\dot{X}^\mu \dot{X}^\nu }
\left(m + \phi(X)\right)\, .
\end{align}
A scalar field equation near the horizon, which is analogous to the electromagnetic case \eqref{eleE}, is
\begin{align}
\partial_r \left(
\sqrt{-\det g} \, g^{rr}  \partial_r \phi
\right)= 0 \, .
\label{phiE}
\end{align}
Near the horizon this has a generic solution which is a logarithmic function
\begin{align}
\phi = m c \log (x /\tilde{c}) \,,
\end{align}
with an arbitrary constant $c$ and a positive constant $\tilde{c}$.
Using this scalar field solution, the effective potential
is written as
\begin{align}
 V_{\rm eff} = \sqrt{\alpha_f x} \left(1  + c \log (x/\tilde{c})\right) \, .
\end{align}
For negative $c$, the scalar force can overcome the gravitational force
and the particle can escape from the horizon. Thus the effective potential
has a maximum, 
at
\begin{align}
x=x_0 \equiv \tilde{c}\exp(-2-1/c)\, .
\end{align}
We can repeat the analysis to derive the
effective Lagrangian near the local maximum $x=x_0$
of the effective potential. The result is
\begin{align}
 {\cal L} \sim
 -\frac{m c
(\alpha_f/\alpha_g)^{1/4}
 }{(2\kappa x_0)^{3/2}}
 \left[ \dot{x}^2 + \kappa^2 (x-x_0)^2 \right] \, .
\label{Lag2}
\end{align}
This effective action turns out to be of the same form as \eqref{Lag} except for the overall
coefficient. 

So, we conclude that the near-horizon particle motion around the maximum of the potential
is universally given by the surface gravity $\kappa$. 
It is independent of the species of the forces 
and their strengths characterized by the slopes $c$ of the potentials,
and also of the particle mass $m$ and the ratio of the metric components $\alpha_g/\alpha_f$.


\section{Horizon as a nest of chaos.}
\label{sec:nest}

In the previous section we have shown that the relativistic particle
near the horizon, pulled by scalar or electromagnetic forces toward outside,
has an effective action \eqref{Lag} (or equivalently \eqref{Lag2})
at the potential maximum near the horizon.
The particle follows the equation of motion with the inverse harmonic potential, 
\begin{align}
\ddot{x} = \kappa^2 (x-x_0) \, .
\label{EOMp}
\end{align}
%
Solving the equation of motion, we 
find the particle trajectory in the $x$ direction as
\begin{equation}
 x = x_0+A e^{\kappa t} + Be^{-\kappa t} \, . 
\end{equation}

The exponentially growing behavior shows a support of a chaos.
Once a coupling to the other coordinates such as $y$ is restored, 
the integrability of the one-dimensional system is lost, and generically a chaos appears. 
If the potential $V(r)$ is minimized along $y$ at some value of $y$,
that position together with $x=x_0$ is a separatrix of the phase space.
The Lyapunov exponent $\lambda$
of the particle motion 
is generally bounded from above by the behavior about the
separatrix with no perturbation by $y$, 
because of the following reasons: First, the particle motion experiences
regions of the configuration space other than the separatrices, and
the Lyapunov exponent is given by an average throughout the motion, thus the averaging
reduces the Lyapunov exponent.  Second, 
when the energy is not large enough to get closer to the separatrices,
the Lyapunov exponent cannot reach the value at the separatrices.
So we obtain a bound for
the Lyapunov exponent of the chaos
caused by this separatrix as
 \begin{align}
  \lambda
  \leq \kappa \, .
\label{kappaL}
 \end{align}

Remarkably, the bound is purely 
given by the surface gravity of the black hole horizon. 
This is because the 
Lagrangian \eqref{Lag} of the particle near the horizon 
does not depend on
the parameters of the system $c$, $m$ and $\alpha_g/\alpha_f$, since they
appears just in the overall factor. The motion of the 
particle is governed only by the surface gravity $\kappa$.
Therefore, the upper bound of the Lyapunov exponent is given
solely by the surface gravity.

\begin{figure*}
\includegraphics[width=5.5cm]{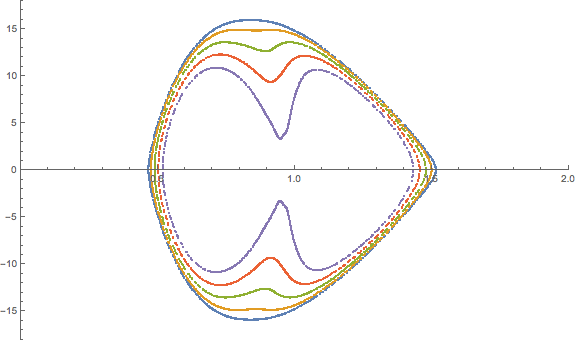}
\hspace{2mm}
\includegraphics[width=5.5cm]{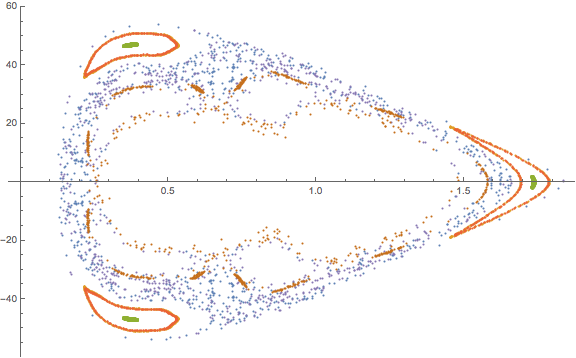}
\hspace{2mm}
\includegraphics[width=5.5cm]{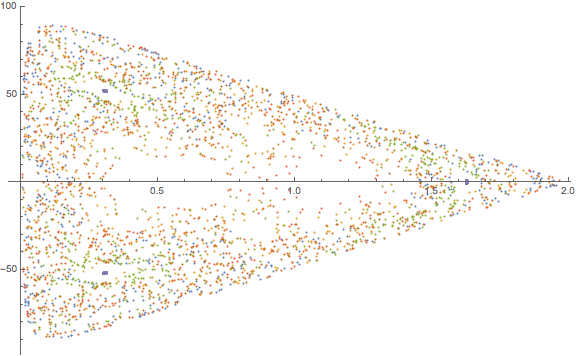}
\caption{(Color online) The Poincar\'e sections of the particle motion near the black hole horizon.
As the energy increases, the KAM tori tend to break.
The energies are $E=15$, $E=40$ and $E=49.5$ respectively, 
from left to right.
}
\label{fig:model2}
\end{figure*}

Using the Hawking temperature of the black hole $T_\text{BH}=\hbar \kappa/2\pi$, we can express
this bound as
\begin{align}
 \lambda
 \leq \frac{2\pi T_\text{BH}}{\hbar}\, .
\label{Mal}
\end{align}
This inequality coincides with the conjectured  chaos bound by Maldacena, Shenker and Stanford
\cite{Maldacena:2015waa}.

Let us briefly discuss  explicit examples. A typical case is a
Schwarzschild black hole in $d$ spacetime dimensions, whose metric is given by
\begin{equation}
  f(r) = 1 - \frac{ r_g{}^{d-3} }{r^{d-3}} \, ,
\end{equation}
where $r_g$ is the horizon radius. 
The mass $M$ and the surface gravity $\kappa$ of this black hole are given by
\begin{equation}
 M = \frac{(d-2)\Omega_{d-2} r_g{}^{d-3}}{16\pi G} \, ,
  \quad
  \kappa
  = \frac{d-3}{2r_g} \, .
\end{equation}
When we derive the Lagrangian \eqref{Lag}, 
first we make an approximation of the near-horizon, which is ensured if
\begin{align}
x \ll r_g \, .
\end{align}
We look at the region around $x\sim x_0$, thus we need to require
$\kappa/2c^2 \ll r_g$, which amounts to 
\begin{align}
\kappa \ll c \, .
\end{align}
This means that the external potential $V(r)$ is set rather strong
compared to the gravitational force. This is natural since
we want to extract the behavior of the particle near the horizon,
while the particle needs to be kept outside of the horizon by being
pulled by the external force.

Let us examine the case of an
extreme Reissner-Nordstr\"om black hole in $d$ dimensions.
The metric is given by
\begin{equation}
  f(r) = \left(1 - \frac{r_g{}^{d-3}}{r^{d-3}}\right)^2.
\end{equation}
The near-horizon metric is 
\begin{gather}
 f(r) \sim \left(
 \frac{(d-3)x}{r_g}
 \right)^2
 \equiv
 \left(\frac{x}{\ell}\right)^2,
 \\
 ds^2 = -\Bigl(\frac{x}{\ell}\Bigr)^2 dt^2 + \Bigl(\frac{\ell}{x}\Bigr)^2 dx^2 + r_g^2 d\Omega^2
 \, ,
\end{gather}
where $\ell$ is the curvature scale of the AdS$_2$ spacetime in the $(t,r)$ part of the metric.
For this metric, the Lagrangian of a point particle is written as
\begin{align}
 {\cal L} &=
 - m\sqrt{
\Bigl(\frac{x}{\ell}\Bigr)^2 - \Bigl(\frac{\ell}{x}\Bigr)^2 \dot x^2 - \dot y^2
 } - V
\notag \\
&\simeq
 -\frac{mx}{\ell}\left[
1 - \frac12 \left(\frac{\ell}{x}\right)^4 \dot x^2
 \right] - V(x) \, ,
\end{align}
hence the effective potential in this case is given by
\begin{equation}
 V_\text{eff} = \frac{x}{\ell} + \frac{V(x)}{m} \, .
\end{equation}
The electrostatic potential $V$ is again a linear function near the horizon
due to \eqref{eleE}, 
thus, there is no stationary point for $V_{\rm eff}$.
This means that there is no extremum which could form a separatrix, thus 
there is no origin of the chaos. The Lyapunov exponent is expected to vanish
$\lambda = 0$
unless
there exists some other source of a chaos. One could say that
$\lambda = 0$
is consistent with
the bound \eqref{Mal} through
the fact the the extreme black hole has zero temperature.


\section{Numerical analyses.}
\label{sec:num}

To illustrate that the black hole horizon is a nest of chaos, 
in this section we
perform numerical analyses of the particle motion, to evaluate
Poincar\'e sections and Lyapunov exponents.
It will be shown that as the motion approaches
the horizon, in other words, the separatrix, a chaos emerges.

Our particle motion near the horizon does not refer to any concrete form of the potential $V(r)$.
So it is instructive to work with the simplest example, a harmonic oscillator in $x$ and $y$ 
space. This example is particularly intriguing since without the gravitational effect the system
is completely integrable and regular, thus it is easier to detect the effect of the black hole horizon.
The Lagrangian of the model we consider is given by
\begin{align}
{\cal L} = -\sqrt{f-\dot{x}^2/f-\dot{y}^2} - \frac{\omega^2}{2}\left((x-x_c)^2+y^2\right) 
\end{align}
with $f\equiv 2\kappa x$, where $\kappa$ is the surface gravity. 
The horizon is located at $x=0$, while the center of the harmonic
potential is off the horizon and located at $(x,y)=(x_c,0)$ with $x_c>0$. 
The harmonic potential has a parameter $\omega$,
but the chaos bound is expected not to depend on this $\omega$. 

We present two kinds of chaos analyses, one is the Poincar\'e section and the other is the Lyapunov exponent. 
The results for the Poincar\'e sections are
shown in Fig.~\ref{fig:model2}. 
The horizontal axis is $x$ while the vertical axis is $p_x$, the conjugate momentum for $x$.
The sections are the slice defined by $y(t)=0$ and $\dot{y}(t)>0$. Three figures are for 
$E=15, 45$ and $49.5$, and we have chosen $\kappa=1/2$, $\omega=10$, $m=1$ and $x_c=1$ for this 
numerical simulation. Colors correspond to different initial conditions at the fixed energy.
Note that the energy has an upper bound because a particle with too much energy will fall into
the black hole horizon and thus chaos is not well-defined.
 
We see from Fig.~\ref{fig:model2} that the Poincar\'e section consists of regular tori for small
energy. But for a larger energy, the motion approaches the horizon $x=0$
and the tori are broken to be transformed to scattered plots which indicate a chaos.

\begin{figure}[hthp]
  \centering
\includegraphics[width=8cm]{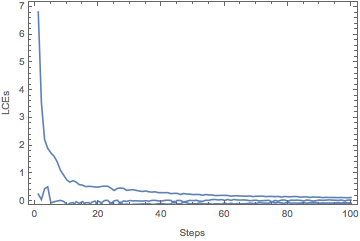}
\caption{(Color online) Numerical analysis of the Lyapunov exponents. 
The horizontal axis is discrete time steps for numerical calculations, whose value is equal to $t$.
The vertical axis is the value of the Lyapunov (characteristic) exponents.
After time steps,
the Lyapunov exponents converge to a set of values.}
\label{fig:model2-L}
 \end{figure}

In Fig.~\ref{fig:model2-L}, we show a numerical calculation of a Lyapunov exponent using the numerical method of Ref.~\cite{Sandri}.
The initial condition is chosen as $y=0.1$ and $\dot{x}=\dot{y}=0$ with $E=49.99$.
The value of the energy is chosen such that the particle can come close enough to
the horizon. There are four Lyapunov exponents as the phase space is four-dimensional,
and because this is a Hamiltonian system, two of the Lyapunov exponents vanish.
Our numerical simulation shows that the four Lyapunov exponents approach 
$(0.08, 0.00, 0.00, -0.08)$ respectively. The largest Lyapunov exponent, whose
numerical value is $\sim 0.08$ in this case, is found to be smaller than the conjectured 
upper bound given by the value $\kappa=1/2$.

In general, Lyapunov exponent depends on the initial condition. So, our choice of the initial condition
is not generic, and is not expected to saturate the bound $\kappa=1/2$. Since we
observed that the obtained value of the Lyapunov exponent is lower than the bound,
it is consistent with our conjectured bound.\footnote{Precisely speaking, 
we cannot eliminate a possibility
that some other nonlinearity in the model causes a different kind of chaos. Here we just demonstrated
a numerical calculation of the simplest model.}

\section{Angular motion of a particle.}
\label{sec:ang}


In this paper, we so far described motion of a relativistic particle near a black hole horizon,
pulled by external forces such as an electromagnetic force or a scalar force.
On the other hand, common force which have been studied in many literature is a centrifugal
force of a neutral particle, that is, an angular momentum.
In fact, the angular momentum creates a similar separatrix which is a homoclinic orbit at
an unstable maximum of the effective potential,
which have been intensively studied \cite{Bombelli:1994ota,Levin:2008yp,PerezGiz:2008yq,Hackmann:2009nh}.
In this section, we show that the angular momentum is not sufficient to bring the motion
of the particle closer to the horizon. Therefore, the angular momentum only is not appropriate 
for our purpose to
probe the black hole horizon.

Let us start with a Lagrangian of a neutral particle, 
\begin{align}
{\cal L}= -m\sqrt{f(r)-\frac{1}{f(r)}\dot{r}^2-r^2\dot{\theta}^2} \, . 
\end{align}
For simplicity here again we consider a spherically symmetric black hole background.
The conserved angular momentum is
\begin{align}
L_\theta = \frac{\partial {\cal L}}{\partial \dot{\theta}} = \frac{mr^2 \dot{\theta}}{\sqrt{f(r)-\frac{1}{f(r)}\dot{r}^2-r^2\dot{\theta}^2}} \, .
\end{align}
The conjugate momentum for $r$ is given by
\begin{align}
P_r = \frac{\partial {\cal L}}{\partial \dot{r}} = m \frac{m\dot{r}/f(r)}{\sqrt{f(r)-\frac{1}{f(r)}\dot{r}^2-r^2\dot{\theta}^2}} \, .
\end{align}
We can define the Hamiltonian as
\begin{align}
{\cal H} \equiv P_r \dot{r} + L_\theta \dot{\theta} - {\cal L} \, .
\end{align}
After an explicit calculation, we obtain
\begin{equation}
 {\cal H} = 
\sqrt{
f \left(
m^2 + f P_r^2 +\frac{L_\theta^2}{r^2} 
\right)
} \, .
\end{equation}
For a very small $P_r$, we can expand this Hamiltonian as
\begin{align}
{\cal H} =
\sqrt{f\left(m^2 + \frac{L_\theta^2}{r^2}\right)}
\left[
1 + 
\frac{f}{m^2 + \frac{L_\theta^2}{r^2}}
\frac{P_r^2}{2} + \cdots
\right] \, .
\end{align}
This means that the effective potential for a very slow motion of the particle is
\begin{align}
V = \sqrt{f(r)} \sqrt{m^2 + \frac{L_\theta^2}{r^2}} \, .
\label{Vang}
\end{align}
The gravitational potential $\sqrt{f(r)}$ is an increasing function of $r$ and
it vanishes at the horizon $r=r_g$. The centrifugal force which is given by the term $L_\theta^2/r^2$
works as a repulsive force. For a sufficiently large value of the angular momentum,
the potential $V$ has an unstable local maximum. The maximum is in fact a homoclinic
circular orbit.

We show that the angular momentum is not sufficient for bringing the particle closer to the black hole
horizon. 
For simplicity,
we consider a Schwarzschild black hole as the background, for which $f(r)=1-r_g/r$.
We substitute 
this to \eqref{Vang}, and consider a limit $L_\theta \to\infty$  to make the effect of the angular 
momentum largest. Then
\begin{align}
 V = \sqrt{1-\frac{r_g}{r}}\,  \frac{L_\theta}{r} \, .
\end{align}
This potential has an unstable maximum at 
\begin{align}
 r = \frac32 r_g \, .
\end{align}
On the other hand the horizon is located at $r=r_g$, so, the unstable orbit is not close to the horizon,
whatever the value of the angular momentum is.

We can show that the same conclusion follows when the background is a Reissner-Nordstr\"om black hole, for which $f(r)=(1-r_+)(1-r_-)$ and the position of the unstable maximum is at $r\geq (3/2)r_g$.

In our previous examples with the electromagnetic force or the scalar force, the 
near-horizon condition
$x (\equiv r-r_g) \ll r_g$ was met. This time, for any large angular momentum, we cannot
reach this region of the near horizon geometry. Therefore, angular motion of a relativistic particle
around the black hole does not have the universal upper bound for the Lyapunov exponent $\lambda = \kappa$.


\section{Generic potential and higher spin forces.}
\label{App:generic}

So far we considered the electromagnetic or the scalar forces. But in general
the force could be of some other origin. Furthermore, depending on the functional
form of the potential, the local maximum of the effective potential may not be near
the horizon.\footnote{One of the peculiar examples was provided
in appendix~\ref{App:scalar} and at Eq.~(\ref{LWyman}), 
where we observed that the Lyapunov exponent is characterized by $\kappa_0$ defined by Eq.~(\ref{defkappa0}) which is the local acceleration
due to the background geometry
at the potential maximum measured by a distant observer.}

In this section, we analyze 
a generic case with a generic potential, which includes
the case when the potential maximum is not close to a black hole horizon.
We examine if the Lyapunov exponent shows a similar universality even in such a case.
In particular, we discuss the case of a hypothetical higher spin
forces, and show that the Lyapunov exponent is again dictated by the surface
gravity $\kappa$, though violating the bound \eqref{MSS}. The violation
could signal some inconsistency of the higher spin fields.

We consider the general spherically-symmetric spacetime given by Eq.~(\ref{smet}), 
and define the local acceleration $\kappa_0(r)$ by
\begin{equation}
 \kappa_0(r)
\equiv
\sqrt{\frac{-g_{tt}}{g_{rr}}}
\frac{d}{dr}\left(\log\sqrt{-g_{tt}}\right)
=
\frac{f'(r)}{2} \sqrt{\frac{g(r)}{f(r)}} ~.
\end{equation}
Following the derivation in Sec.~\ref{sec:PM}, we find that the Lagrangian of a particle moving in the radial direction is given by
\begin{equation}
 {\cal L} =
 -m \sqrt{f(r) - \frac{\dot r^2}{g(r)}} - V(r)
 \simeq
 m\left[
\frac{\dot r^2}{2\sqrt{f}g} - V_\text{eff}(r)
 \right], 
\end{equation}
where
\begin{equation}
 V_\text{eff}(r) \equiv \sqrt{f(r)}  + m^{-1} V(r) \, .
\end{equation}
At a local maximum of $V_\text{eff}(r)$, at which $V'_\text{eff}(r\equiv r_0)=0$,
Lagrangian is approximated as
\begin{align}
 {\cal L} &=
 \frac{m}{2\sqrt{f(r_0)} g(r_0)}\dot r^2 - m\left[
 V_\text{eff}(r_0)
 + \frac12 V_\text{eff}''(r_0) (r-r_0)^2
 \right]
 \notag \\
 &\simeq
 \frac{m}{2\sqrt{f(r_0)}g(r_0)}
 \left[
 \dot r^2 +
 \tilde \kappa^2 (r-r_0)^2
 \right],
\end{align}
where we have neglected the constant part in the potential, and
\begin{equation}
 \tilde \kappa^2 \equiv
  \kappa_0{}^2(r_0)
  -\frac12 g(r_0)\left(
 f''(r_0) + 2m^{-1}\sqrt{f(r_0)} V''(r_0)
 \right).
 \label{kappat}
\end{equation}
Using this $\tilde\kappa$, the Lyapunov exponent is given by
\begin{align}
\lambda = \tilde\kappa \, .
\end{align}
This is a formula for generic potential $V$.

In view of Eq.~\eqref{kappat}, 
the condition that the Lyapunov exponent for the particle trajectory equals to the local acceleration $\kappa_0$ is that the terms involving $f''(r_0)$ and $V''(r_0)$ can be neglected in $\tilde \kappa$ defined above.\footnote{
The cases discussed in the main text and appendix~\ref{App:scalar} are examples of such a situation.}

The electromagnetic potential we argued 
in Sec.~\ref{sec:PM} is linear around the horizon, as a consequence of the
electromagnetic equation of motion. If the equation of motion for the force field is different,
we need to adopt a different functional form of the potential $V$.
Here we demonstrate the case of higher spin forces. Surprisingly, we 
find that the upper bound of the Lyapunov exponent is again dictated by
the surface gravity $\kappa$, but the coefficient grows with the spin.

The massless higher spin potential $A_{00\cdots 0}$ will obey an equation of motion in the background of
the black hole,
\begin{align}
\partial_r 
\left(\sqrt{-\det g} \, g^{rr} \left(g^{00}\right)^s \partial_r A_{00\cdots 0} \right)= 0 \, .
\end{align}
Here $s$ is the spin of the field $A$ which has $s$ indices. 
This equation is an analogue of the electromagnetic equation \eqref{eleE}.
Since $g_{rr}\sim 1/x$ and $g_{00}\sim x$, a generic vacuum solution
of the equation is
\begin{align}
A_{00\cdots 0} = c_1 x^s + c_2 \, ,
\end{align}
where $c_1$ and $c_2$ are constant parameters.
A reparameterization-invariant coupling to the charged particle is
\begin{align}
V = e \left[
\frac{dX^{\mu_1}}{dt} \cdots \frac{dX^{\mu_s}}{dt}
A_{\mu_1 \cdots \mu_s} 
\right]^{1/s}. 
\end{align}
where $e$ is a charge.
When only the component $A_{0\cdots 0}$ is nonzero, the worldline
gauge fixing $X^0=t$ leads to a potential term
\begin{align}
V= e \left(A_{0\cdots 0}\right)^{1/s} \, .
\label{VA}
\end{align}
For the spin 1 case of the electromagnetic force, it reproduces the linear potential.

For a generic case\footnote{For the special case $c_2 = 0$, the potential
\eqref{VA} is equivalent to that of the electromagnetic case, thus the 
Lyapunov exponent is bounded by $\kappa$.}
 $c_2\neq 0$, 
near the horizon $x\sim 0$ we can expand \eqref{VA} as
\begin{align}
V = \mbox{const.} + \frac{ec_1}{s (c_2)^{1-1/s}} x^s + \cdots.
\label{Vxs}
\end{align}
The power of the potential is given by the spin $s$. For this potential, in \eqref{kappat}
the term $V''$ does not vanish, thus the Lyapunov exponent differs from the universal value $\kappa$. In fact, using \eqref{Vxs}, for the spherically symmetric black holes
with $\beta_f=\beta_g=1$,  
we find 
\begin{align}
\lambda = \sqrt{2s-1} \, \kappa \, .
\end{align}
First, interestingly, again the Lyapunov exponent is given by the surface gravity $\kappa$.
Second, the coefficient grows as $\sqrt{2s-1}$, and it violates the bound \eqref{MSS}.
In the electromagnetic case $s=1$ it reduces to $\lambda = \kappa$.

The growth of the Lyapunov exponent for the higher spin force can be compared with
the CFT calculation studied in \cite{Roberts:2014ifa,Perlmutter:2016pkf}, where it was
shown that the Lyapunov exponent of the out-of-time-ordered correlator is
given by $\lambda = (s-1)2\pi T$ which violates the bound for $s>1$. The $s$-dependence
is different from ours, because in our case we introduced the higher spin field 
only for pulling the particle off the horizon. The coincidence that both Lyapunov exponents
violate the bound for $s>1$ is interesting, and it suggests
that higher spin theories are described by a unitary CFT only in a special situation where
the Lyapunov exponent cancels off by some mechanism.


\section{Conclusion and discussions.}

In this paper, we discovered that the chaos of a particle probing the black hole horizon
has a universal upper bound for the Lyapunov exponent. The upper bound is given by
the surface gravity of the horizon, \eqref{kappaL}, and in particular is independent of
the forces which pull the particle not to fall into the horizon. The separatrix is located
very near the horizon. We note that it is not realized for common situations of an orbiting particle.

The derived upper bound of the Lyapunov exponent coincides with the one given 
by quite a generic argument by Maldacena, Shenker and Stanford \cite{Maldacena:2015waa}.
The argument was originally based on shock waves near black hole horizons 
\cite{Shenker:2013pqa,Shenker:2013yza} and
the AdS/CFT correspondence. Here we demonstrated that the same upper bound
can be obtained for particles near the horizon.

According to the AdS/CFT dictionary, a particle near the black hole horizon 
is dual to some composite operator in a dual CFT. 
If the particle is an approximation of a low energy fundamental string,
the finite-size effect of an elongated worldsheet near the horizon needs to be considered,
which can alter the Lyapunov exponent analysis itself.\footnote{%
Chaotic motion of fundamental strings in various AdS-like geometries have been studied by
\cite{Zayas:2010fs} -- \cite{Chervonyi:2015ima}.}
Supposing on the other hand that 
the particle is an approximation of a D-particle or a wrapped D-brane,
our model may be mapped naturally to string theory.
In any case, extension to include larger worldvolume for the chaos analysis
is important (see \cite{Hashimoto:2016wme} for a chaotic D7-brane motion).

Since our bound is dictated by the surface gravity, we may argue that our universal chaos 
originates in redshift near the black hole horizon. In fact, the redshift means a very slow motion
of a particle as measured from an observer at the spatial infinity.
In appendix \ref{sec:Red}, we analyzed a simple model with redshift and
found 
that a chaos emerges even in this case, although
this result itself may not be enough to assure that the origin of the chaos is the redshift.
For the case of the shock waves \cite{Shenker:2013pqa,Shenker:2013yza} the redshift is in fact the origin
of the universal chaos. 
Furthermore, as shown in section \ref{sec:ang}, simply orbiting particles do not come 
closer to the horizon, but it has a separatrix. It is consistent with an observation by Polchinski
\cite{Polchinski:2015cea} concerning Ruelle resonances. In summary, the relation between the
separatrix and the redshift could be not so simple, and  further analysis of various separatrices
will be necessary to reveal the reason of the universality.

Finally, in view of the AdS/CFT correspondence, a gauge-theory interpretation
of our separatrix is important and needs to be studied. Some weak-coupling analysis of chaotic D0-brane
motion \cite{Aref'eva:1997es,Aref'eva:1998mk,Asplund:2011qj,Asplund:2012tg,Aoki:2015uha,Asano:2015eha,Gur-Ari:2015rcq,Berkowitz:2016znt} 
(Yang-Mills models were originally investigated in \cite{Matinyan:1981dj,Savvidy:1982jk}) 
will show the separatrix of our type, if exists. Then 
the mystery of the black hole horizons,
such as the information loss problem \cite{Hawking:2015qqa,Hawking:2016msc},
may be revealed, since the interpretation
of the infinite redshift of a classical gravity in terms of strongly coupled gauge theories
or quantum information could be provided there. Further exploration of the
universal chaos will benefit physics of black holes, gauge theories and quantum information.


\acknowledgments
K.H.\ would like to thank valuable discussions with M.~Hanada, N.~Iizuka, M.~Kimura, K.~Murata, A.~Tseytlin, B.~Yoshida and K.~Yoshida.
The work of K.H.\ was supported in part by JSPS KAKENHI Grant Numbers 15H03658, 15K13483.
The work of N.T.\ was supported in part by JSPS KAKENHI Grant Number 16H06932.

\appendix

\section{Surface gravity of a spherically-symmetric static black hole.}
\label{App:surfacegravity}

Consider a black hole with metric given by
\begin{equation}
 ds^2 = g_{tt}(r) dt^2 + g_{rr}(r) dr^2 + \cdots.
\end{equation}
The surface gravity of this black hole is calculated as follows.
The unit tangent vector $v^\mu$ of the world line of a static particle, that is, the four-velocity of the particle is given by
\begin{equation}
 v^\mu =
  \left(
   g_{tt}{}^{-1/2},0,0,\ldots
  \right).
\end{equation}
Then, the acceleration of this trajectory $a^\mu$ is calculated as
\begin{equation}
 a^\mu = v^\nu \nabla_\nu v^\mu.
\end{equation}
We can show that the components of this vector vanish other than the $r$ component, which is given by
\begin{equation}
 a^r = \Gamma^r_{tt}\left(v^t\right)^2
  =
  - \frac{1}{g_{rr}} \frac{\partial}{\partial r}
  \left(
   \log\sqrt{- g_{tt}}
   \right).
\end{equation}
From this expression, the proper acceleration $a$ of this particle is obtained as
\begin{equation}
 a \equiv \sqrt{a^\mu a_\mu} 
  =
  \frac{1}{\sqrt{g_{rr}}} \frac{d}{dr} \left(\log\sqrt{-g_{tt}}\right).
\end{equation}
Also, the redshift factor $V$ at the particle location is calculated from the timelike Killing vector $\left(\partial_t\right)^\mu$ as
\begin{equation}
 V^2 = - \left(\partial_t\right)^2 = -g_{tt} \, .
\end{equation}
Using the quantities obtained above, the surface gravity $\kappa$ may be expressed as
\begin{equation}
 \kappa = Va\Bigr|_\text{horizon}
  =
  \left. \sqrt{\frac{-g_{tt}}{g_{rr}}}
   \frac{d}{dr} \left( \log\sqrt{-g_{tt}} \right)
  \right|_\text{horizon}
  .
\end{equation}

In any dimensions, the temperature of a black hole is related to the surface gravity by
\begin{equation}
 kT=\frac{\hbar \kappa}{2\pi c} \, .
\end{equation}

 \begin{figure*}[htbp]
  \centering
\includegraphics[width=5.5cm]{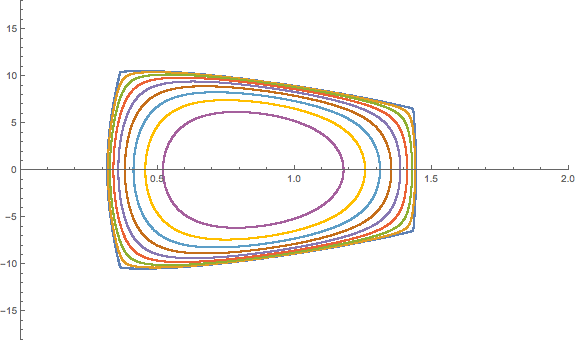}
\hspace{1mm}
\includegraphics[width=5.5cm]{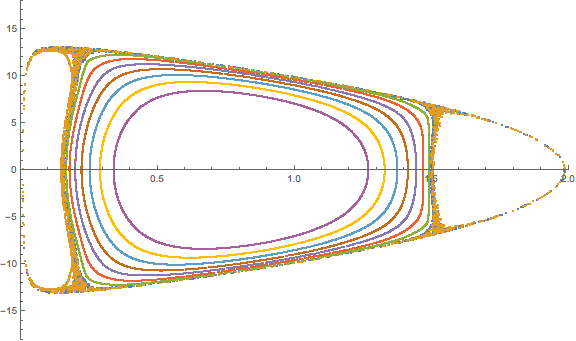}
\hspace{1mm}
\includegraphics[width=5.5cm]{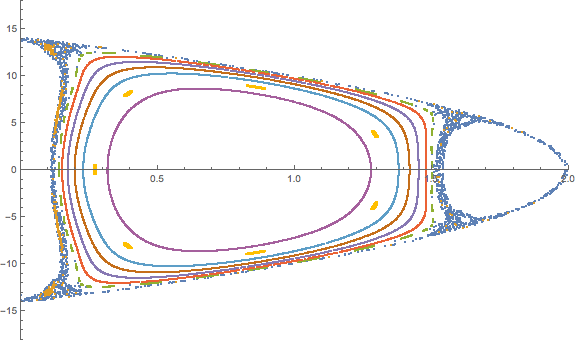}
\caption{(Color online) The Poincar\'e sections of the model \eqref{model2}. The sections are the $(x,p_x)$-plane
with $y=0$ and $\dot{y}>0$. The energies are $E=40$, $E=49$ and $E=49.99$ from left to right. For larger energies,
the particle motion comes closer to the line $x=0$ where the KAM tori start to break to form a scattered plot,
and a chaos appears. For the numerical simulation, we have chosen $x_c=1$ and $\omega =10$.}
\label{fig:model}
 \end{figure*}

\section{Redshift and chaos.}
\label{sec:Red}

We have worked on the motion of a relativistic particle \eqref{Lag}, and found
a separatrix which provides a nest of chaos. The separatrix is an inverted harmonic oscillator,
and its existence is based on the black hole horizon. At the separatrix, the motion
of the particle becomes extremely slow, and this is the physical reason of the
emergent chaos: Exponentially slow motion enlarges the difference in phase space motion
at later time. Around the black hole horizon, there is a physical reason to expect the
slow motion. Near the horizon, the factor $g_{rr}$ diverges, and the motion is
infinitely redshifted, to realize the slow motion. So, it is quite natural to expect
that a chaos always appear at black hole horizons.

In this appendix, we present a simple model and discuss a
possible relation between the redshift and the separatrix. 
Let us consider the following non-relativistic model with a ``horizon'',
\begin{align}
{\cal L} = \frac{g(x,y)}{2} \left(\dot{x}^2+\dot{y}^2\right) - \frac{\omega^2}{2}\left((x-x_c)^2+y^2\right).
\label{model2}
\end{align}
Here $x(t)$ and $y(t)$ specify the location of a particle in a two-dimensional plane, 
and we put a harmonic oscillator potential centered at $(x,y)=(x_c,0)$ for illustrating the originally-regular system,
as before.
The presence of a black hole ``horizon'' is given by the ``metric''
\begin{align}
g(x,y) \equiv \frac{1}{x} \, ,
\end{align}
which diverges at $x=0$. The ``horizon'' is located at a line $x=0$, and the motion is
restricted to the region $x>0$.

As expected, a numerical analysis of 
this simple model exhibits a chaos when the trajectory approaches the horizon.
In Fig.~\ref{fig:model}, we show Poincar\'e sections $(x,p_x)$ defined by the condition
$y=0$ and $\dot{y}>0$. We have chosen $x_c=1$ and $\omega=10$ for our numerical simulations. 
From left to right in Fig.~\ref{fig:model}, we have chosen $E=40, E=49$ and $E=49.99$ respectively.
When the energy $E$ approaches  $E=50$, the trajectory approaches the horizon. From the figures,
we find that some KAM tori collapse and we have a chaos. The collapsed trajectories
show up only for values of the energy close to $E=50$, 
that is, only for trajectories getting close to the
``horizon'' $x=0$.

Of course, if we took $g=1$ instead, this model would be integrable and there is no chaos. Therefore, the origin of the
chaos of this model is the ``horizon''. It can also be seen from the Poincar\'e sections, because
a particle which feels the redshift at the horizon is shown to have a scattered section and so is subject to a chaos.

To discuss a relationship between the simple redshift model here and the separatrix which we have 
studied in this paper, let us expand the model \eqref{model2} around the ``horizon'' $x=0$.
We obtain
\begin{align}
{\cal L} = \frac{1}{2x} \dot{x}^2 + \omega^2 x_c \; x \, .
\label{Lxx}
\end{align}
The potential term is regular, while the kinetic term has the divergence at $x=0$.
Now, we make a coordinate redefinition for $x\geq 0$, 
\begin{align}
X = 2\sqrt{x} \, .
\label{sre}
\end{align}
Then the Lagrangian \eqref{Lxx} is brought to the form
\begin{align}
{\cal L} = \frac{1}{2} \dot{X}^2 + \frac{\omega^2 x_c}{4}X^2 \, .
\end{align}
This is nothing but a model with an inverted harmonic oscillator potential, and is a typical
situation near the separatrix. 
The slow motion was realized by the redshift in the original model, but after the coordinate redefinition,
the physics is shown to be identical to an inverted harmonic
potential for $X$.
In this manner, the redshift can be understood as an origin of a chaos.

Note that in this model
the separatrix is on top of the ``horizon'' $x=0$, while  
in the particle model studied in section \ref{sec:PM} the separatrix is a little bit separated from the
horizon. So in general these two models are not equivalent to each other. It would be interesting if 
some more concrete physical origin of the universal chaos found in this paper can be discussed
from the viewpoint of the redshift.

%
%

\section{Particle motion with scalar force.}
\label{App:scalar}

In Sec.~\ref{sec:PM}, we analyzed the motion of a point particle with scalar interaction, and found that its motion is governed by the surface gravity of the background black hole. One of caveats in this analysis was that such a static scalar field diverges at the horizon and the spacetime becomes singular once the gravitational backreaction is taken into account. In this appendix, we repeat the analysis taking an exact solution of Einstein equations with scalar field as the background and observe how it changes the conclusion derived in the main text.

The background solution we take in this appendix is
the exact solution of a spherically-symmetric static object with scalar charge,
the Wyman solution given by~\cite{Wyman:1981bd}
\begin{align}
ds^2 &=
 -
 f^\gamma(r) dt^2
 +
 f^{-\gamma}(r) dr^2
 +
r^2 f^{1 - \gamma}(r)  d\Omega^2 \, ,
\label{Wmetric}
\\
\Phi &=
 \frac{q}{b\sqrt{4\pi G}}\log f(r) \, ,
\label{Wscalar}
\end{align}
where
\begin{equation}
 f(r) \equiv 1 - \frac{b}{r},
  \quad
  b = 2 \sqrt{m^2 + q^2} \geq 2m,
  \quad
  \gamma = \frac{2m}{b}\leq 1.
  \label{Wdef}
\end{equation}
This solution is known to be equivalent to the Janis-Newman-Winicour solution~\cite{Janis:1968zz},
and the relationship between them is addressed in, e.g., Ref.~\cite{Virbhadra:1997ie}.
The equalities in Eq.~(\ref{Wdef}) hold when the scalar field vanishes ($q=0$).
This is a solution of the Einstein equation with a massless scalar field:
\begin{align}
&G_{\mu\nu} = 8\pi G\left(
\nabla_\mu \Phi \nabla_\nu \Phi 
- \frac12 \nabla_\rho \Phi \nabla^\rho \Phi g_{\mu\nu}
\right),
\notag \\
 &\nabla_\mu \nabla^\nu \Phi = 0 \, .
\label{EeqW}
\end{align}
This spacetime has a naked singularity at $r=b$, where the circumferential radius $\sqrt{g_{\theta\theta}}$ becomes zero.
Since the horizon is absent in this spacetime, it does not make sense to calculate the surface gravity.
For example, the
quantity corresponding to the 
surface gravity~(\ref{surfacegravity}) at $x=b$ is evaluated as
\begin{equation}
\kappa =
 \sqrt{\frac{-g_{tt}}{g_{rr}}}\frac{d}{dr}\left(\log \sqrt{-g_{tt}}\right)\biggr|_{r=b}
 =
\frac{\gamma}{2}\frac{x^{\gamma-1}}{b^\gamma}
\biggr|_{x=0}~,
\end{equation}
where we introduced $x \equiv r - b$.
Because $\gamma-1\leq 0$, it diverges unless $\gamma=1$, that is, when the scalar field vanishes.
Instead of this $\kappa$, in the following we will focus on the quantity defined by
\begin{equation}
 \frac{\gamma}{2}\frac{x_0{}^{\gamma-1}}{b^\gamma} \equiv \kappa_0 \, ,
  \label{defkappa0}
\end{equation}
which may be regarded as the proper acceleration generated by the background geometry
at $x = x_0$ measured by an observer at infinity.
Also, we assume that the scalar potential felt by a point particle is given by
\begin{equation}
\phi = c \log \frac{x}{\tilde c}\,.
\label{phiofx}
\end{equation}
It is different from $\Phi$ in Eq.~(\ref{Wscalar}) by a constant shift, which is free to add since Eq.~(\ref{EeqW}) depend only on derivatives of $\Phi$, and also by an overall coefficient, which corresponds to the coupling constant between the scalar field and the particle.

The Lagrangian for the radial motion in this case is given by
\begin{align}
{\cal L} &=
- \sqrt{f^\gamma - f^{-\gamma}\dot x^2}
\left[
m + m c \log \left(\frac{x}{\tilde c}\right)
\right]
\notag \\
 &\simeq
 m \left( 1 + c \log \frac{x}{\tilde c} \right)
 \left[
 f^{-\frac{3\gamma}{2}}
\dot x^2
- f^{\frac{\gamma}{2}}
\right],
\label{Lscalar}
\end{align}
from which we define the effective potential by
\begin{equation}
V_\text{eff} \equiv \sqrt f \left(1 + c \log \frac{x}{\tilde c}\right).
\end{equation}
The approximations used at ``$\simeq$'' is valid when $\gamma\sim 1$, $\tilde c /b \ll 1$ and $\dot x^2$ is sufficiently small.
One of the characterization of this regime is that
$g_{tt}$ and $g_{rr}^{-1}$ becomes sufficiently small:
\begin{equation}
1 - \frac{b}{r} \simeq \frac{x_0}{b} = \frac{\tilde c}{b}e^{-\frac1c - \frac2\gamma}\ll 1 \, .
\end{equation}

Assuming $c<0$, we find that the maximum of $V_\text{eff}$ is located at 
\begin{equation}
x = 
\tilde c 
\exp\left(
-\frac1c - \frac2\gamma 
\right)
\equiv x_0
\, .
\label{x0scalar}
\end{equation}
We find that the Lagrangian is approximated around $x= x_0$ as
\begin{equation}
{\cal L} =
-\frac{mce^{\frac{3}{2c} + \frac{3}{\gamma}}\sqrt\gamma}{2\sqrt2 \bigl(\tilde c \kappa_0\bigr)^{3/2}}
\left[
\dot x^2 + \kappa_0{}^2 (x-x_0)^2
	\right]\, .
\label{LWyman}
\end{equation}
This equation implies that the Lyapunov exponent in this case is given by $\lambda = \kappa_0$, which is independent of $c$.
In this sense, the Lyapunov exponent of the particle trajectory shows the universality similar to that found in Sec.~\ref{sec:PM} even in this case, if we focus on the local acceleration $\kappa_0$.

In Sec.~\ref{sec:PM}, we assumed that the background geometry is fixed to that of a spherically-symmetric static black hole,
and the scalar hair is treated as a probe field on it.
Relationship of this case and that addressed in this appendix is subtle,
but it may be fair to suppose that the case of Sec.~\ref{sec:PM} is reproduced from the above derivation
if the background geometry at $x=x_0$ is sufficiently close to a usual black hole geometry.
To realize such a situation,
$g_{\theta\theta}$ of Eq.~(\ref{Wmetric}) must be sufficiently close to $r^2$ when $x = x_0$, that is,
\begin{equation}
\left(1-\frac{b}{r}\right)^{1-\gamma} r^2 \simeq r^2
\qquad
(\text{at $x =x_0$}) \, .
\end{equation}
When this condition is satisfied, the metric around $x=x_0$ is well approximated by the Schwarzschild metric.
At $x = x_0$, the factor in the above equation is evaluated as
\begin{equation}
\left(1-\frac{b}{r}\right)^{1-\gamma}\biggr|_{x=x_0}
=
e^{-\left(\frac1c+\frac2\gamma\right)(1-\gamma)}
\left(\frac{\tilde c r}{b^2}\right)^{\gamma-1}
,
\end{equation}
which must be sufficiently close to the unity.
For this to be satisfied, since $r\simeq b$ we just need
\begin{equation}
\left(\frac{\tilde c}{b}\right)^{\gamma-1}
= \exp\left[
(1-\gamma)\log \frac{\tilde c}{b}
\right]
\simeq 1
\, .
\end{equation}
This is satisfied if
\begin{equation}
\left|
(1-\gamma)\log \frac{\tilde c}{b}
\right|\ll 1 \, .
\end{equation}
Since $\log(\tilde c / b)$ takes a large value when $\tilde c/ b\ll 1$, $\gamma$
must be very close to 1 to suppress it.


\end{document}